\documentclass[english,aps,prl,reprint,superscriptaddress,nolongbibliography]{revtex4-2}
\usepackage[T1]{fontenc}
\usepackage[latin9]{inputenc}

\usepackage{amsmath}
\usepackage{amssymb}
\usepackage{graphicx}
\usepackage{wasysym}
\usepackage{xcolor}
\usepackage{physics}
\usepackage{xr}

\makeatletter
\usepackage{pslatex}
\usepackage[space]{grffile}

\usepackage{babel}

\makeatother

\begin{document}

\title{Photon Cycling and Laser Cooling of an Asymmetric Top Molecule}

% \author{Grace K. Li}
% \email{kehui\_li@g.harvard.edu}
% \author{Christian Hallas}
% \author{John M. Doyle}
% \affiliation{Department of Physics, Harvard University, Cambridge, MA 02138, USA}
% \affiliation{Harvard-MIT Center for Ultracold Atoms, Cambridge, MA 02138, USA}

\author{Grace K. Li}
\email{kehui\_li@g.harvard.edu}
\affiliation{Department of Physics, Harvard University, Cambridge, MA 02138, USA}
\affiliation{Harvard-MIT Center for Ultracold Atoms, Cambridge, MA 02138, USA}

\author{Giseok Lee}
\affiliation{Department of Physics, Harvard University, Cambridge, MA 02138, USA}
\affiliation{Harvard-MIT Center for Ultracold Atoms, Cambridge, MA 02138, USA}

\author{Jack Mango}
\affiliation{Department of Physics, Harvard University, Cambridge, MA 02138, USA}
\affiliation{Harvard-MIT Center for Ultracold Atoms, Cambridge, MA 02138, USA}

\author{Hana Lampson}
\affiliation{Department of Physics, Harvard University, Cambridge, MA 02138, USA}
\affiliation{Harvard-MIT Center for Ultracold Atoms, Cambridge, MA 02138, USA}

\author{YongWoong Lee}
\affiliation{Department of Physics, Korea University, 145 Anam-ro, Seongbuk-gu, Seoul, 02841, Republic of Korea}

\author{Winston Wang}
\affiliation{Department of Physics, Harvard University, Cambridge, MA 02138, USA}
\affiliation{Harvard-MIT Center for Ultracold Atoms, Cambridge, MA 02138, USA}

\author{Avikar Periwal}
\affiliation{Department of Physics, Harvard University, Cambridge, MA 02138, USA}
\affiliation{Harvard-MIT Center for Ultracold Atoms, Cambridge, MA 02138, USA}
\affiliation{Department of Physics, Massachusetts Institute of Technology, Cambridge, MA 02139, USA}

\author{Nathaniel B. Vilas}
\affiliation{Department of Physics, University of California, Berkeley, CA 94720, USA
}

\author{Alexander Frenett}
\affiliation{Facility for Rare Isotope Beams, Michigan State University, East Lansing, MI 48824, USA}

\author{Lo\"{i}c Anderegg}
\affiliation{Department of Physics and Astronomy, University of Southern California, Los Angeles, CA 90089, USA}

\author{John M. Doyle}

\affiliation{Department of Physics, Harvard University, Cambridge, MA 02138, USA}
\affiliation{Harvard-MIT Center for Ultracold Atoms, Cambridge, MA 02138, USA}

\date{\today}

\date{\today}

\begin{abstract}
We realize two-dimensional magnetically-assisted Sisyphus laser cooling of an asymmetric top molecule (ATM), calcium monoamide (CaNH$_2$). Vibrational state closure is achieved with $41.1 \pm 6.3$ photons scatters using optical pumping of the $\Tilde{X}[3_1]$ state. Photon-cycling measurements show good agreement with branching ratios determined by dispersed fluorescence spectroscopy. Rotational closure is maintained by driving the $\Tilde{X}[1_{11}] \to \Tilde{A} [0_{00}]$ transition. The observed absence of additional state leakage channels broadens the scope of molecular laser cooling to include ATMs, which are the most general geometric class of molecules and possess the richest internal structure. Future applications of quantum controlled ATMs include new quantum information platforms and searches for physics beyond the Standard Model.
\end{abstract}

\maketitle
Since its invention, laser cooling has become a cornerstone technique in AMO physics and quantum information processing. Continued advances have enabled full quantum control of atoms in conservative potentials, including optical traps and tweezer arrays,
% ~\cite{greiner2002quantum,endres2016atom,barredo2016atom,anderegg2019optical}
leading to applications in quantum computation~\cite{graham2022multi, bluvstein2024logical}, quantum simulation~\cite{luciuk2016evidence,ebadi2021quantum,saint2019dynamical}, quantum sensing~\cite{marciniak2022optimal}, optical clocks~\cite{marshall2025high,aeppli2024clock}, and precision measurements~\cite{parker2018measurement,parker2015first, kozyryev2021enhanced}. Originally realized in alkali atoms and alkaline-earth ions, laser cooling has since been extended to a diverse range of species, including alkaline-earth atoms~\cite{katori1999optimal}, lanthanides~\cite{kuwamoto1999magneto,lu2011strongly}, transition metals~\cite{uhlenberg2000magneto,griesmaier2005bose}, exotic atoms~\cite{baker2021laser,gloggler2024positronium}, and molecules~\cite{zhelyazkova2014laser,shuman2010laser,collopy20183d,padilla2025magneto,vilas2022magneto,lasner2025magneto}.

Molecules have a vastly larger Hilbert space than atoms, offering new opportunities for quantum science through their rich ro-vibrational structure, which has energy scales spanning many orders of magnitude. This molecular complexity has advanced quantum information science (QIS)~\cite{kaufman2021quantum,bao2023dipolar,holland2023demand} and searches for physics beyond the Standard Model (BSM)~\cite{mitra2022quantum,jansen2014perspective}. 
% Early breakthroughs in the production and control of ultracold molecules have since evolved into two major experimental approaches: assembly from laser-cooled atoms, primarily for bialkali molecules~\cite{ni2008high,danzl2008quantum}, and direct laser cooling, which 
% Direct laser cooling has been demonstrated for diatomic~\cite{zhelyazkova2014laser,shuman2010laser,collopy20183d,padilla2025magneto}, linear triatomic~\cite{vilas2022magneto,lasner2025magneto}, and
% symmetric-top molecules~\cite{mitra2020direct}. However, asymmetric-top molecules (ATMs), which constitute the vast majority of molecular species and offer some of the richest opportunities in quantum science, have thus far remained beyond the reach of full laser cooling and quantum control.
Recent advances in the quantum control of linear molecules have highlighted parity-doublet states -- which are generic to polyatomic molecules -- as promising for BSM searches and QIS due to their long coherence times, spin-1 structure, and strong low-field polarizability~\cite{anderegg2023quantum, robichaud2026parity}. In a linear molecule, the coherence time of the $l$-based parity doublets is ultimately limited by the radiative lifetime of the excited bending mode. In contrast, in an asymmetric top molecule (ATM), parity-doublet states exist in the vibrational ground state, resulting in extremely long coherence times. 

The scientific potential of ATMs, which are the most general and abundant class of molecules, has
%for quantum information encoding~\cite{albert2020robust,hong2026universal}, precision measurements~\cite{mitra2022quantum,jansen2014perspective,augenbraun2020molecular}, and high-precision spectroscopy relevant to interstellar chemistry and astrophysics~\cite{kleiner2019spectroscopy}, this 
generated significant interest in achieving full quantum control and cooling to the ultracold regime. Spectroscopic studies and theory proposals have identified several candidate ATMs for optical cycling and laser cooling~\cite{frenett2024vibrational, augenbraun2020molecular}, including functionalized arenes such as CaOPh~\cite{zhu2022functionalizing,mitra2022pathway,augenbraun2022high}. However, experimental efforts to optically cycle CaOPh observed scattering of only two photons, far below the 
%$\sim20$ photons
number predicted from spectroscopic measurements~\cite{burchesky2023engineered}. The origin of this discrepancy remains unclear, leaving open the question of whether the structure of ATMs is fundamentally incompatible with efficient optical cycling, or if another mechanism played the determinative role in suppressing cycling~\cite{wojcik2026unraveling}.

In this work, we report the optical cycling and laser cooling of an asymmetric top molecule, calcium monoamide (CaNH$_2$). To quantify photon cycling, we perform a precision beam-deflection measurement and observe scattering of $41.1 \pm 6.3$ photons, in agreement with predictions from quantitative dispersed-fluorescence spectroscopy. We also demonstrate two-dimensional, magnetically assisted Sisyphus cooling to reduce the transverse temperature of a molecular beam of CaNH$_2$ from $12$ mK to $1.4$ mK. These results establish the feasibility of full quantum control and laser cooling of asymmetric-top molecules and pave the way toward magneto-optical trapping and deep laser cooling of this class of molecules.

\begin{figure*}[t]
    \centering
    \includegraphics[width=0.95\linewidth]{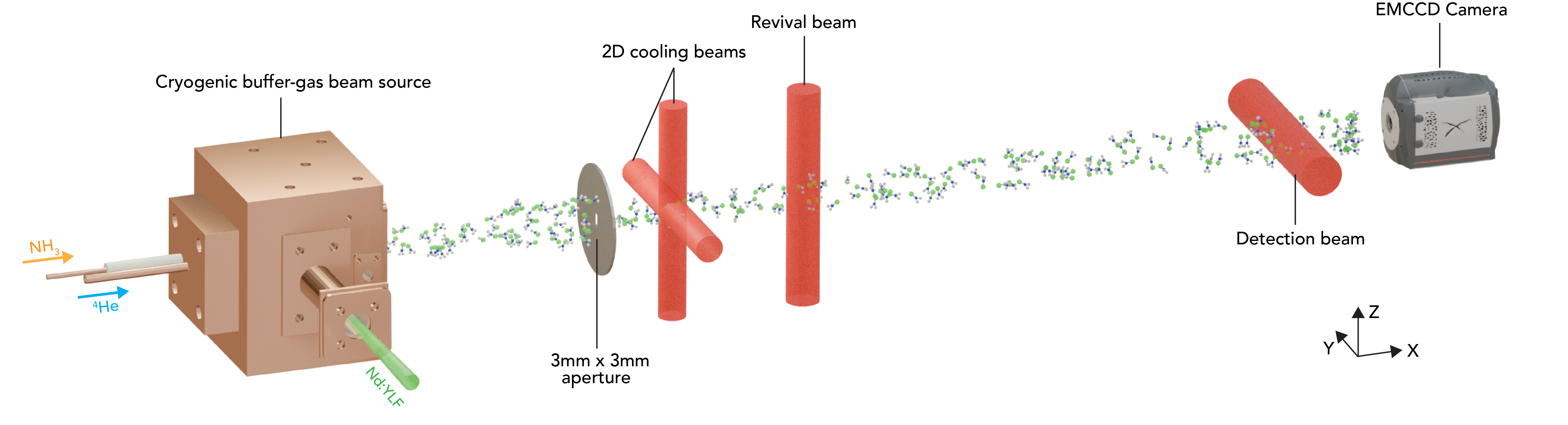}
    \caption{Schematic of the 2D transverse laser cooling experiment. A 657~nm laser enters the buffer gas cell from the opposite of the ablation laser, to drive the $4s^2\,{}^1S_0 \rightarrow 4s4p\,{}^3P_1$ transition in calcium for an enhanced chemical reaction rate to CaNH$_2$. The distance between the exit aperture of the buffer gas cell and the collimating aperture is $\sim$ 40 cm. The cooling beams are $\sim$ 10 cm downstream from the aperture, followed by the revival beams. $\sim$ 100 cm downstream from the cooling beams, an expanded horizontal beam is used for detection. The EMCCD camera that images the beam is positioned at the end of the beam line.}
    \label{fig:CoolingSetup}
\end{figure*}

CaNH$_2$ is a planar, near-prolate asymmetric top molecule with $C_{2v}$ symmetry. A rotationally closed transition exists from the electronic ground state $\Tilde{X}^2 A_1$ to the first excited state $\Tilde{A}^2 B_2$, between rotational states $1_{11}$ ($N = 1$, $k_a = 1$, $k_c= 1$) and $0_{00}$ ($N'=0$, $k_a'=0$, $k_c'=0$)~\cite{augenbraun2020molecular,frenett2024vibrational}. Due to the representation of $\Tilde{X}$ and $\Tilde{A}$ under the $C_{2v}$ group, any transition between the two must be b-type ($\Delta k_a = \pm 1$, $\Delta k_c = \pm 1$). As a result of these selection rules, $\Tilde{A}^2 B_2 \text{ } [0_{00}]$ can only decay back into $\Tilde{X}^2 A_1 \text{ } [1_{11}]$. The vibrational branching from the $\Tilde{A}$ state has been measured at high resolution using dispersed fluorescence spectroscopy~\cite{jack2026mango}, and the Franck-Condon factor of the diagonal decay is estimated to be $\sim 0.95$. The dominant vibrational branching channel is into the $3_1$ state, which has one excitation in the Ca-N stretching mode, with branching ratio $0.03$. The stretching mode has $A_1$ representation under $C_{2v}$, which is the same as the vibrational ground state $0_0$, and thus the same rotational selection rules apply to this decay, and no rotational branching is generally expected. In the current experiment, we demonstrate laser cooling and deflection with only a $3_1$ vibrational repump. Higher vibrational leakage channels have been observed and quantified with dispersed laser-induced fluorescence spectroscopy~\cite{jack2026mango}, and can be repumped with additional lasers.

\begin{figure}[h]
    \centering
    \includegraphics[width=1\linewidth]{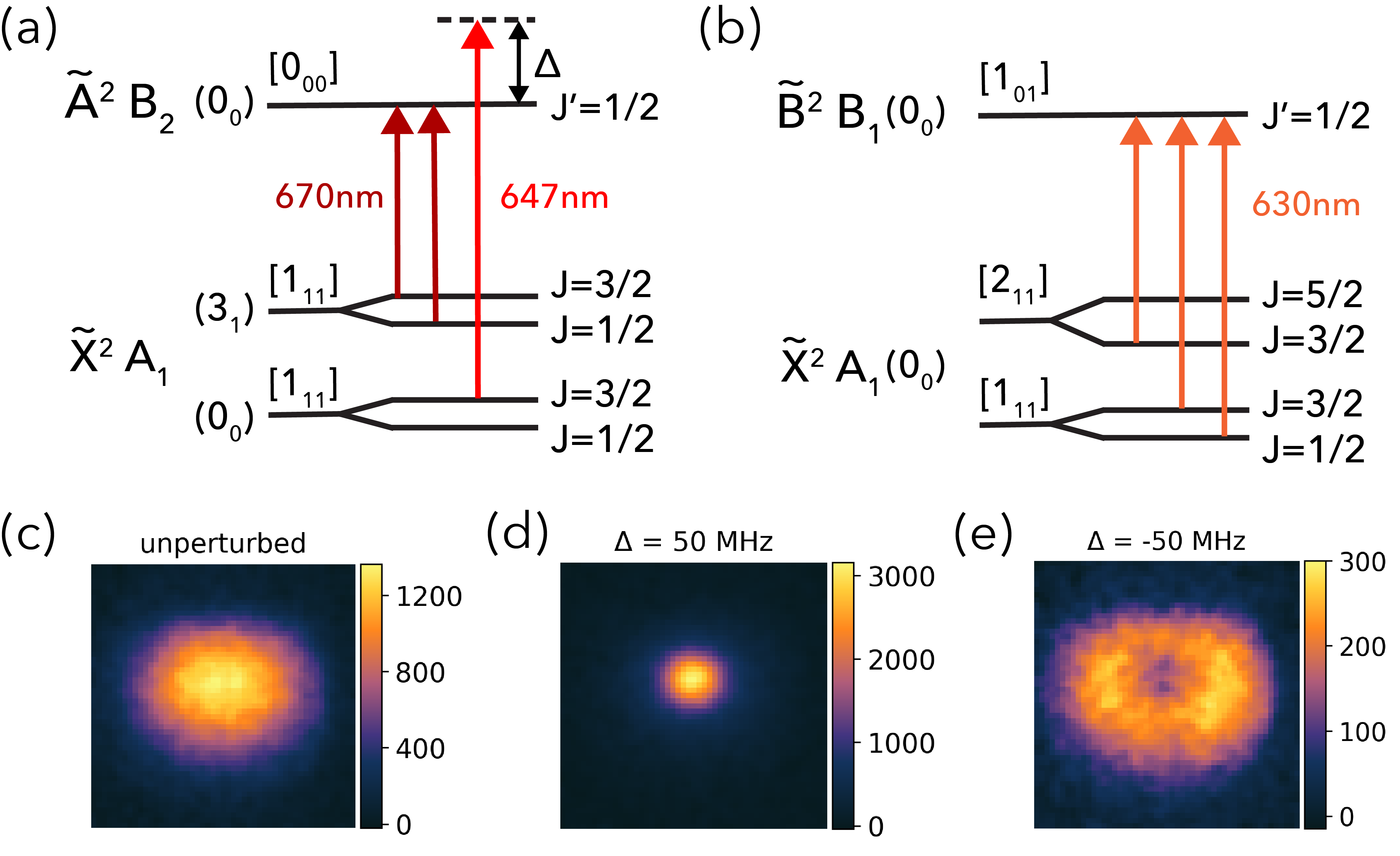}
    \caption{(a) Structure of the cycling transition of CaNH$_2$, in vibrational ground state and $3_1$, as well as the lasers frequencies used for the transverse cooling. The spin-rotation splitting between $J=1/2$ and $J=3/2$ in both $0_0$ and $3_1$ is 65~MHz. For clarity, vibrational state labels are given in round brackets, and rotational states in square brackets. (b) Transitions used for imaging. The second row shows camera images of the molecule beam, (c) without any laser interaction, (d) with the crossed laser beams blue-detuned ($\Delta = $ 50 MHz), and (e) with the crossed laser beams red-detuned ($\Delta = $-50 MHz).}
    \label{fig:LevelsAndPhotos}
\end{figure}

The experimental setup is shown in Figure~\ref{fig:CoolingSetup}. CaNH$_2$ molecules are produced in a cryogenic buffer gas cell. Calcium atoms are released by ablating a metallic calcium target with an Nd:YLF laser, and ammonia gas is flowed into the cell to form CaNH$_2$. We observe a significant enhancement to the reaction rate by exciting the calcium atoms to the metastable $4s4p\,{}^3P_1$ state~\cite{bopegedera1987laser,jadbabaie2020enhanced}. A beam of CaNH$_2$ molecules exits the cell with a forward velocity of $230\pm30$\,m/s. A 3$\times$3\,mm aperture collimates the molecule beam. Downstream from the aperture, two pairs of retro-reflected laser beams intersect the molecule beam to provide horizontal and vertical cooling. The power in the horizontal and vertical beams are equal and the $1/e^2$ beam diameter is 4~mm. All four lasers beams are linearly polarized along the X axis. The frequency configuration of these lasers is shown in Figure \ref{fig:LevelsAndPhotos}(a). Three pairs of Helmoltz coils, one along each spatial axis, are positioned around the cooling region to generate a magnetic field. After the cooling region, the molecule beam passes through a clean-up region, where only the $3_1$ repump is present to optically pump molecules into $0_0$ for detection. Finally, a horizontal laser beam images the molecule beam onto an EMCCD camera that points in the $-X$ direction, 95~cm downstream from the cooling region, in order to observe changes in the density distribution along both of the cooling axes. The $\Tilde{X}\,^2A_1 \rightarrow \Tilde{B}\,^2B_1$ transitions are used for imaging because they are spectrally separated from the cooling transitions, allowing scattered cooling light to be filtered. The imaging frequencies are shown in Fig.~\ref{fig:LevelsAndPhotos}(b). Molecules are detected on the $\Tilde{X}^2A_1(0_0)[1_{11}] \rightarrow \Tilde{B}^2B_1(0_0)[1_{01}],J=1/2$ transitions, while the $\Tilde{X}^2A_1(0_0)[2_{11}],J=3/2 \rightarrow \Tilde{B}^2B_1(0_0)[1_{01}],J=1/2$ transition is driven for rotational closure.

An averaged image of the unperturbed molecule beam is shown in Figure \ref{fig:LevelsAndPhotos}(c). Note that the apparent smaller width in the vertical direction is due to the finite size of the imaging beam. The $1/e^2$ diameter of the molecule beam fits to $24.2 \pm 1.6$ mm. Taking into account the forward velocity distribution of the molecule beam, we estimate the temperature to be $11.6 \pm 0.4$ mK through Monte-Carlo simulations (Supplement I). Figure \ref{fig:LevelsAndPhotos}(d) shows an averaged image of the molecules with 0.4 W laser power in each cooling direction and a detuning of $\Delta = 50$ MHz, $B_y=2$\,G. The transverse temperature of the cooled molecules fits to $1.4 \pm 0.1$ mK. In Figure \ref{fig:LevelsAndPhotos}(e) we show an averaged image of the molecule beam with red detuning ($\Delta = -50$ MHz), which leads to heating. As expected, the molecules are pushed away from zero velocity, resulting in a donut-shaped distribution.

\begin{figure}
    \centering
    \includegraphics[width=0.95\linewidth]{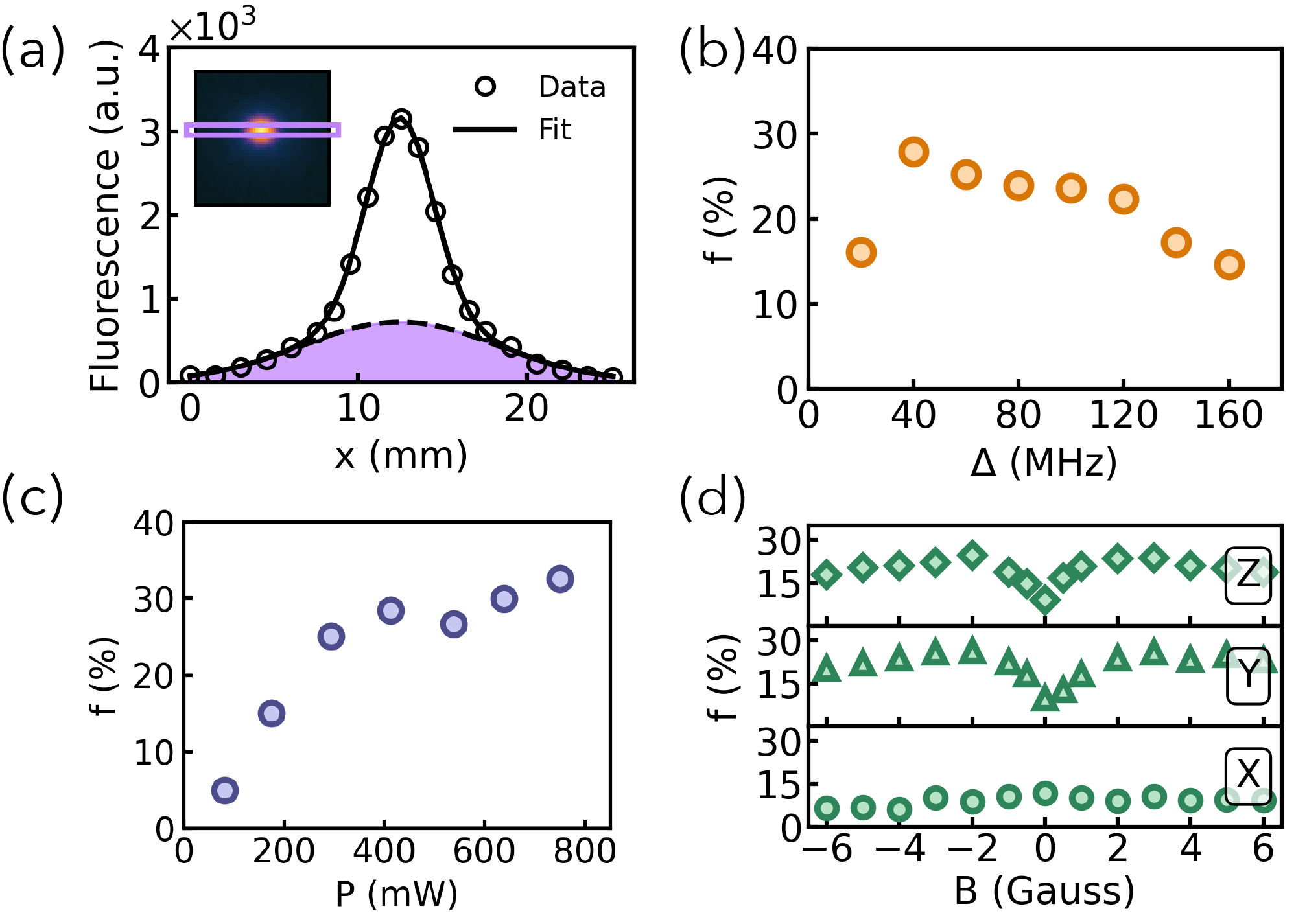}
    \caption{(a) Example of a 1D slice of the image and 2D fit. (b) The cooled fraction as a function of laser detuning. (c) Cooled fraction as a function of total power in the cooling laser. (d) Cooled fraction as functions of magnetic field along three spatial axes. $Z$ and $Y$ are transverse to the polarization of the lasers, $X$ is parallel.}
    \label{fig:scans}
\end{figure}

To further confirm that the observed effect is magnetically assisted Sisyphus cooling, and to characterize the cooling force under different experimental conditions, we study the cooling efficiency as a function of laser detuning, laser power, and magnetic field. To quantify the cooling efficiency, we fit the camera images to the sum of two two-dimensional Gaussian functions and define the ``cooled fraction" (f) as the ratio of the volume under the narrower 2D Gaussian to the total volume (Figure~\ref{fig:scans}(a)). This metric, which reflects the capture velocity, is used in our subsequent analysis. Another figure of merit is the final temperature, which is determined from the width of the narrow Gaussian through Monte Carlo simulations (Supplement I). We find that this width varies little across the explored parameter space, and therefore report only the temperature obtained under optimal cooling conditions, which is $1.4 \pm 0.1$ mK.

\begin{figure*}
    \centering
    \includegraphics[width=0.9\linewidth]{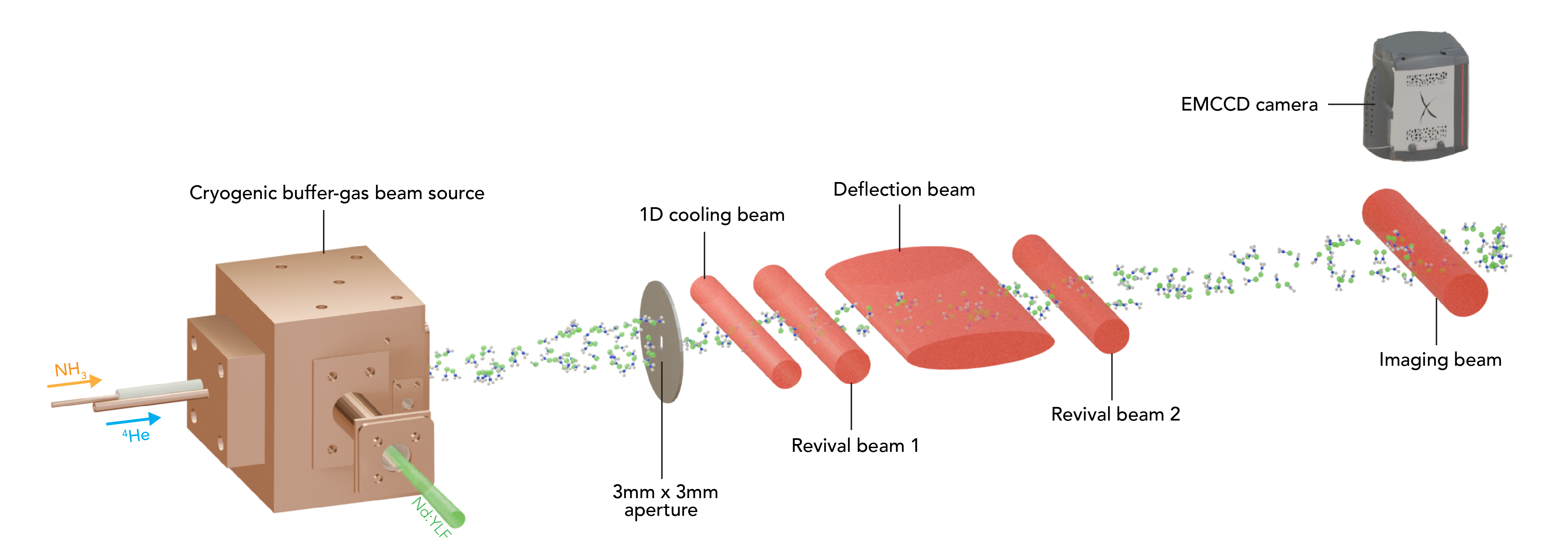}
    \caption{Schematic of the deflection experiment. The buffer gas cell and collimating aperture remain unchanged from the 2D cooling experiment. The horizontal cooling beam is $\sim$ 10 cm downstream from the aperture, followed by the first revival beam. A horizontally expanded deflection beam is positioned $\sim$ 5 cm downstream from the cooling beam, followed by the second revival beam. $\sim$95 cm downstream from the deflection beam, a horizontal imaging beam is used for inducing fluorescence.}
    \label{fig:DeflectionSetup}
\end{figure*}

The cooled fraction as a function of laser detuning is shown in Figure~\ref{fig:scans}(b), where the detuning is defined relative to the $J=3/2$ levels of $\Tilde{X}^2A_1$ $(0_0)$. With 0.4~W of optical power in each direction, the cooling feature (double-Gaussian profile) is observed for $\Delta$ > 0, and the cooling fraction is optimal around $\Delta$ = 40 MHz. Note that at the optimal detuning, the laser is red-detuned to the $J=1/2$ levels, suggesting that cooling effect relies primarily on interaction between the light field and the $J=3/2$ states. The cooled fraction as a function of total laser power is shown in Figure~\ref{fig:scans}(c) where, as expected, higher optical power leads to higher capture velocity, hence higher cooled fraction.

A prominent feature of magnetically-assisted Sisyphus cooling is that the dark state remixing relies on a transverse magnetic field~\cite{emile1993magnetically}. Thus with zero B-field or B-field co-aligned with the light polarization, the cooling effect should vanish. To test this, we align the polarization of both cooling beams along the $X$ axis, which points along the molecule beam. The magnetic field scan along all three spatial axes is shown in Figure~\ref{fig:scans}. We see that with zero B-field, the cooling fraction is very low, but non-zero --- possibly from imperfect cancellation of earth's magnetic field. As the magnetic fields in the transverse directions ($Y$ and $Z$) are increased from zero, the cooling fraction increases from $10\%$ to about $30\%$, before slowly dropping at higher fields, where rapid remixing starts to disrupt the Sisyphus cooling process. In contrast, when $B_x$ is increased, the cooling fraction remained low, consistent with the conditions of magnetically-assisted Sisyphus effect.
 
\begin{figure}[t]
    \centering
    \includegraphics[width=0.95\linewidth]{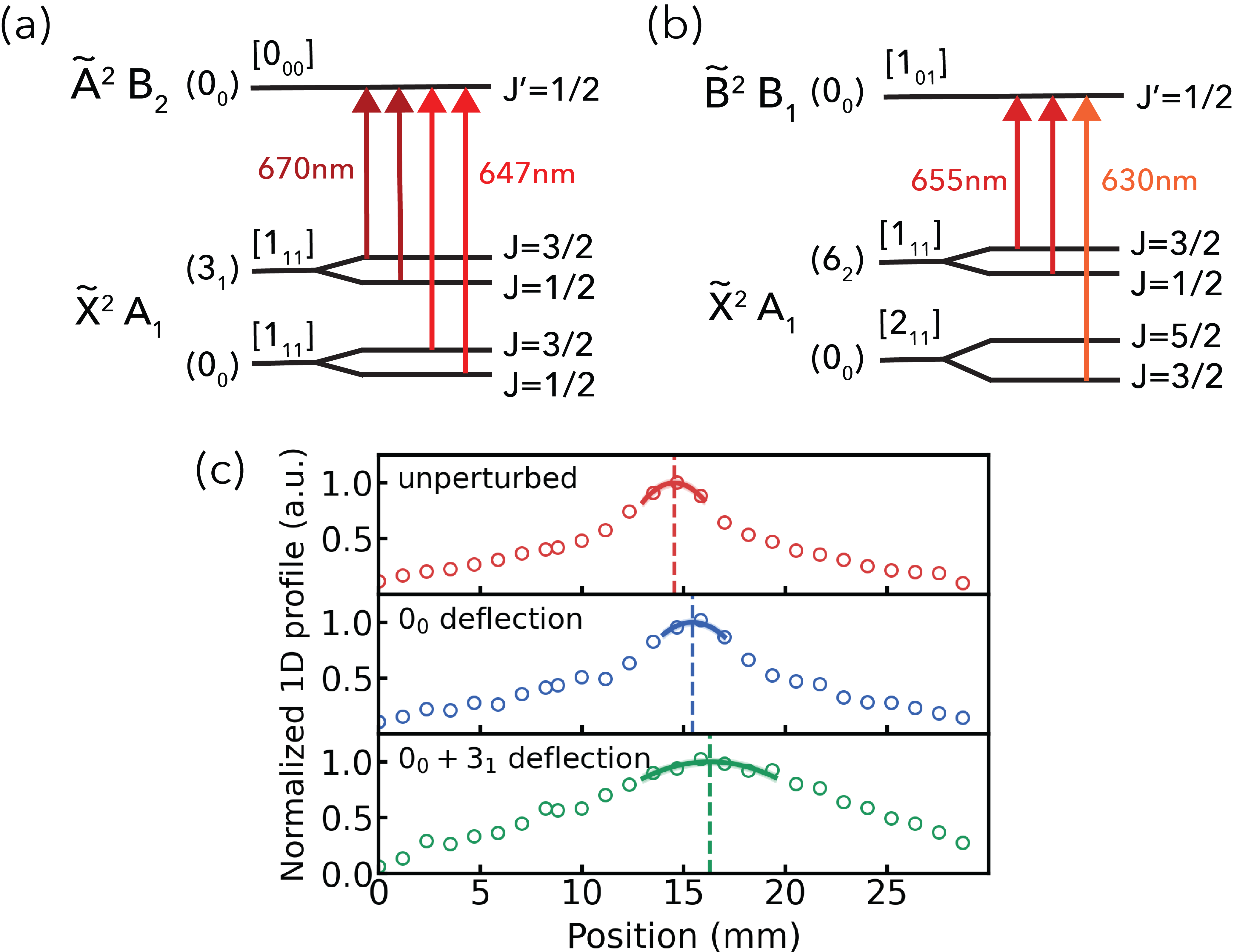}
    \caption{ Beam deflection level structure and data. (a) Optical transitions driven by the deflection light. (b) Additional transitions used to repump the $6_2$ state. (c) One-dimensional sum of the camera images showing the undeflected beam (top), deflection using the $0_0$ cycling transition (middle), and deflection using the $0_0$ cycling transition together with the $3_1$ vibrational repump (bottom).}
    \label{fig:DeflectionResults}
\end{figure}

While the two-dimensional cooling experiment proves the feasibility of transverse cooling of CaNH$_2$, three-dimensional laser cooling and trapping, which is typically preceded by laser slowing, requires the ability to scatter $\sim 10^4$ photons. To quantitatively assess photon cycling in this asymmetric top molecule, we perform a beam-deflection experiment and determine the number of photons scattered by the molecules (Figure \ref{fig:DeflectionSetup}). We report deflection measurements under two conditions: first, by driving only the main cycling transition, $\Tilde{X}^2A_1 (0_0)[1_{11}] \rightarrow \Tilde{A}^2B_2 (0_0)[0_{00}],J=1/2$; and second, with the addition of a vibrational repump addressing $\Tilde{X}^2A_1 (3_1)[1_{11}] \rightarrow \Tilde{A}^2B_2 (0_0)[0_{00}], J=1/2$. Both transitions are shown in Figure~\ref{fig:DeflectionResults}(a). In the second configuration, molecules are depleted from both the $0_0$ and $3_1$ states during the deflection process, necessitating an additional vibrational repump for fluorescence imaging. The next most significant vibrational decay channel is the $6_2$ state, which shares the same representation as $0_0$ and can, in principle, be driven to the same excited state without rotational branching. However, due to limitations in laser tunability, population in this state is instead repumped through the $\Tilde{B}$ state via the transition $\Tilde{X}^2A_1(6_2)[1_{11}] \rightarrow \Tilde{B}^2B_1(0_0)[1_{01}], J=1/2$. To maintain rotational closure, an additional repump addresses the $\Tilde{X}^2A_1(0_0)[2_{11}], J=3/2 \rightarrow \Tilde{B}^2B_1(0_0)[1_{01}], J=1/2$ transition, as shown in Figure~\ref{fig:DeflectionResults}(b).

The experimental setup is shown in Fig.~\ref{fig:DeflectionSetup}. As in the two-dimensional cooling experiment, the molecular beam is collimated. After passing through the aperture, the beam is cooled along the horizontal direction to reduce the transverse velocity spread and facilitate observation of the deflection signal. Due to limited available power for the $3_1$ vibrational repump laser, only the cycling transition is used during the one-dimensional cooling stage. The $3_1$ repump laser (as well as the $6_2$ and $(0_0)[2_{11}]$ repumps in the second configuration) is subsequently applied in the first revival beam to return molecules to the $0_0$ state.

%After the first revival region, the molecules interact with a deflection beam resonant with the cycling transition and, in the second configuration, the $3_1$ repumping transition.
After the first revival region, the molecules interact with a deflection beam resonant with the cycling transition. In the second configuration, the deflection beam also includes light resonant with the $3_1$ repumping transition. The beam is elongated along the molecular-beam axis to provide an interaction length of approximately 4 cm. Slightly downstream of the deflection region, molecules optically pumped into vibrational dark states are returned to the cycling manifold by the $3_1$ repump laser, together with the $6_2$ and $(0_0)[2_{11}]$ repumps in the second configuration. The molecules are then imaged using a horizontal probe beam resonant with the $\Tilde{X}\to\Tilde{A}$ cycling transition. The resulting fluorescence is collected by an EMCCD camera mounted vertically, thereby minimizing background noise from any upstream laser scatter.

The deflection results obtained using only the main cycling transition are shown in the second row of Figure~\ref{fig:DeflectionResults}(c). With a laser power of 0.8~W, we observe nearly complete depletion of the $\Tilde{X}(0_0)[1_{11}]$ state. Following revival with the $3_1$ repump in the second revival beam, we measure a beam deflection of $0.90 \pm 0.09$~mm. The average forward velocity of the molecular beam, measured immediately after the deflection experiment, is $240 \pm 30$~m/s, which translates to $\sim 4.0$~ms flight time between the deflection and imaging regions. Thus, the center-of-mass transverse velocity of the molecules changes by $0.23 \pm 0.04$~m/s = $(20.6 \pm 3.3)$ $v_\text{recoil}$, where $v_\text{recoil} = \frac{h}{m\lambda}$ = 0.011 m/s. This is in agreement with Monte Carlo trajectory simulations. We therefore conclude that $20.6\pm 3.3$ photons can be scattered by driving the cycling transition, without any vibrational or rotational repumps, i.e. a $95.1 \pm 0.8\%$ diagonal branching fraction.

The deflection results obtained with the additional $3_1$ vibrational repump are shown in the third row of Figure~\ref{fig:DeflectionResults}(c), where we observe a beam displacement of $1.87 \pm 0.14$ mm, corresponding to a transverse velocity change of $(41.1 \pm 6.3)v_\mathrm{recoil}$. We note that the broadening of the distribution can be attributed to heating due to photon scattering and the addition of uncooled $2_{11}$ natural population by the first revival beam. Owing to the finite interaction length, the deflection beam does not completely deplete the molecular population in the $0_0$ and $3_1$ states. Consequently, the deflection analysis is performed using only the molecules that are optically pumped into the $6_2$ state and subsequently revived for detection. From the measured survival fraction of $12\%$, we infer, using the probability model in Supplement II, a combined branching ratio of $98.47 \pm \mathrm{0.23}\%$ for decays into the $0_0$ and $3_1$ states. This value is consistent with Monte Carlo simulations of the optical cycling and deflection processes and corresponds to a photon budget of $65 \pm \mathrm{10}$. The inferred branching fraction to the $3_1$ state is $3.37 \pm \mathrm{0.83}\%$. The ratio between the measured $3_1$ and $0_0$ branching fractions is consistent with dispersed fluorescence of 0.0336 $\pm$ 0.0007 within experimental uncertainty~\cite{jack2026mango}.

In summary, we demonstrate the feasibility of direct laser cooling and photon cycling in asymmetric top molecules by realizing two-dimensional magnetically assisted Sisyphus cooling and observing beam deflection consistent with the scattering of $41.1 \pm 6.3$ photons using a single cycling transition and one vibrational repump, in agreement with previous spectroscopic studies. These results partially resolve the CaOPh mystery, indicating that the previously observed failure to achieve substantial photon cycling in CaOPh does not arise from an inherent incompatibility between asymmetric top structure and optical cycling. Instead, the limitation may originate from molecule-specific properties, such as a high density of electronic or vibrational states in CaOPh. Our results show that laser-cooling techniques can be extended to ATMs, the broadest and most structurally complex class of molecules. These results lay the groundwork for laser slowing and magneto-optical trapping of CaNH$_2$ and other ATMs~\cite{frenett2024vibrational,vadachkoria2025electronic}, establishing a new platform for ultracold quantum science with applications in precision measurement and searches for physics beyond the Standard Model~\cite{mitra2022quantum,jansen2014perspective,augenbraun2020molecular}.

The authors would like to thank Alireza Eghdamian for assistance with the 3D modeling used in the figures. This work is supported by AOARD, NSF, ARO, QSA, and Heising-Simons Foundation.

\bibliography{references}

\end{document}

\title{Supplemental Material: \emph{Photon Cycling and Laser Cooling of an Asymmetric Top Molecule}}

\author{Grace K. Li}
\email{kehui\_li@g.harvard.edu}
\affiliation{Department of Physics, Harvard University, Cambridge, MA 02138, USA}
\affiliation{Harvard-MIT Center for Ultracold Atoms, Cambridge, MA 02138, USA}

\author{Giseok Lee}
\affiliation{Department of Physics, Harvard University, Cambridge, MA 02138, USA}
\affiliation{Harvard-MIT Center for Ultracold Atoms, Cambridge, MA 02138, USA}

\author{Jack Mango}
\affiliation{Department of Physics, Harvard University, Cambridge, MA 02138, USA}
\affiliation{Harvard-MIT Center for Ultracold Atoms, Cambridge, MA 02138, USA}

\author{Hana Lampson}
\affiliation{Department of Physics, Harvard University, Cambridge, MA 02138, USA}
\affiliation{Harvard-MIT Center for Ultracold Atoms, Cambridge, MA 02138, USA}

\author{YongWoong Lee}
\affiliation{Department of Physics, Korea University, 145 Anam-ro, Seongbuk-gu, Seoul, 02841, Republic of Korea}

\author{Winston Wang}
\affiliation{Department of Physics, Harvard University, Cambridge, MA 02138, USA}
\affiliation{Harvard-MIT Center for Ultracold Atoms, Cambridge, MA 02138, USA}

\author{Avikar Periwal}
\affiliation{Department of Physics, Harvard University, Cambridge, MA 02138, USA}
\affiliation{Harvard-MIT Center for Ultracold Atoms, Cambridge, MA 02138, USA}
\affiliation{Department of Physics, Massachusetts Institute of Technology, Cambridge, MA 02139, USA}

\author{Nathaniel B. Vilas}
\affiliation{Department of Physics, University of California, Berkeley, CA 94720, USA
}

\author{Alexander Frenett}
\affiliation{Facility for Rare Isotope Beams, Michigan State University, East Lansing, MI 48824, USA}

\author{Lo\"{i}c Anderegg}
\affiliation{Department of Physics and Astronomy, University of Southern California, Los Angeles, CA 90089, USA}

\author{John M. Doyle}

\affiliation{Department of Physics, Harvard University, Cambridge, MA 02138, USA}
\affiliation{Harvard-MIT Center for Ultracold Atoms, Cambridge, MA 02138, USA}
\maketitle

\section{Monte Carlo temperature estimates}

We use Monte Carlo simulations to calculate the transverse temperature of our molecular beam both with and without the magnetically assisted Sisyphus cooling. Initial transverse velocities are sampled from a thermal distribution with a temperature of 5.4\,K, matching the approximate operating temperature of the cryogenic cell the molecules are produced in. The longitudinal velocities are sampled from a gaussian distribution, with mean forward velocity of $230$ m/s and a standard deviation of $30$ m/s. Initial positions of the molecules at the cell exit aperture are sampled from a gaussian distribution with standard deviation $\sigma_0$, which is inferred from the experimentally measured unperturbed beam width.

\begin{figure}
    \centering
    \includegraphics[width=0.5\columnwidth]{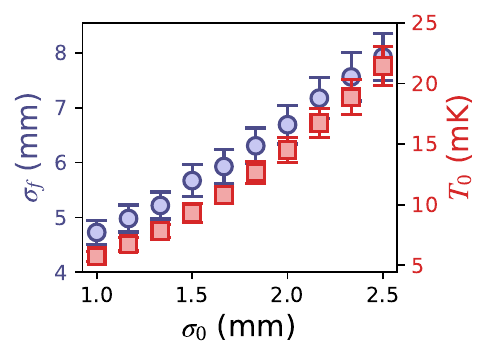}
    \caption{Final simulated beam widths $\sigma_f$ and unperturbed transverse temperatures $T_0$ as a function of effective initial beam radius $\sigma_0$ sampled at the cell aperture.}
    \label{fig:mc}
\end{figure}

Trajectories are simulated for $10^8$ molecules starting at the cell exit aperture using simple kinematics. Two limiting apertures downstream reduce the effective transverse temperature of the beam. First, a 12.7 mm diameter hole in the 4\,K radiation  of the cryogenic buffer gas beam box, then a 3$\times$3 mm square collimating aperture. Molecules that do not pass through these apertures are discarded from the simulations. Finally, the molecules are propagated to the imaging region, where the final position and velocity distributions are recorded. Both distributions are approximately gaussian, with standard deviations of $\sigma_f$ and $\sigma_v$ respectively. To determine the appropriate value of $\sigma_0$, the simulation is repeated multiple times for a range of $\sigma_0$'s, and the resulting $\sigma_f$ is computed for each  (Fig.\ref{fig:mc}). Uncertainties on the predicted $\sigma_f$'s are estimated by randomly varying distances between the cell, apertures and imaging region, matching experimental uncertainties in these distances. The experimentally measured standard deviation of the unperturbed beam is $\sigma_f =$ 6.06 $\pm$ 0.1 mm, therefore we conclude $\sigma_0$ = 1.7 $\pm$ 0.1 mm.

Once $\sigma_0$ is determined, the unperturbed transverse velocity distribution can be simulated using kinematics of ballistic expansion, with constraints given by the limiting apertures. We use a Monte Carlo error-weighted least-squares fit to estimate the width of the final velocity distribution $\sigma_v$, and assign an unperturbed transverse temperature $T_0$ according to
% From the simulations, the relationship between $\sigma_f$ and $\sigma_v^2$ is observed to be roughly linear. A Monte Carlo error-weighted least-squares fit gives the $\sigma_v^2$. Treating the transverse velocity distribution as thermal, we can assign an unperturbed transverse temperature $T_0$ according to
\begin{equation}
    \sigma_v^2 = k_B T_0 / m_{\text{CaNH}_2}
\end{equation}
% as a function of $\sigma_f$. 
% The experimentally observed value of $\sigma_f$ is used with the optimal fit parameters to obtain an estimate for $T_0$, with error propagated from the linear fit and the gaussian fit of the beam size. 
This gives an estimated initial transverse temperature of $11.6 \pm 0.4$ mK.

The temperature of the cooled beam is determined using the same Monte Carlo approach. 
% The value of $\sigma_0$ is randomly varied with a mean value derived from the linear relationship between $\sigma_f$ and $\sigma_0$ in the unperturbed temperature simulations, and standard deviation propagated from the error in $\sigma_0$. 
At the cooling region the velocity distribution for the molecular cloud is resampled for the molecules having a velocity below a capture velocity, $v_c$. The value for $v_c$ is chosen such that the proportion of cooled molecules matches the experimentally observed cooling fraction. For a cooling fraction of $40\%$ this is a velocity of $\sim0.8$ m/s. The resampled velocities are drawn from a thermal distribution with a temperature $T_f$. 
% The relationship between $\sigma_f$ and $T_f$ is observed to be linear. 
Monte Carlo error-weighted least-squares fitting is used to get a relationship between $\sigma_f$ and $T_f$. The experimentally measured standard deviation of the cooled beam is $\sigma_f$ = 1.94 $\pm$ 0.02 mm. With this, we estimate the final temperature to be $T_f$ = 1.4 $\pm$ 0.1 mK.
% and the optimal parameters are used to estimate $T_f$ from the experimentally observed value of $\sigma_f =  $xx $\pm$ xx mm. Error is propagated from the linear fit and the double gaussian fits to the observed molecular beam.

\section{Branching ratio estimates}
In this section we discuss how measured populations after deflection are used to calculate the probability that a molecule is lost to a dark state.
We assume an unknown constant photon scattering rate $\gamma$ for molecules in bright states ($0_0$ and $3_1$). After an interaction time $t$, molecules that remained in the bright state have scattered $N_\text{max}=\gamma t$ photons, and molecules that decayed into a dark state scattered fewer photons.

The number of photons a molecule scatters before decaying into a dark state is given by a geometric distribution. Let the branching ratio into states other than $0_0$ and $3_1$ be $p$. The probability that a molecule becomes dark upon the $n$th photon scatter is 
\begin{align}
    P(n) = p(1-p)^{n-1}
\end{align}
The probability that a molecule remains bright after $N_\text{max}$ scatters is 
\begin{equation}
    \Pmax = (1-p)^{N_\text{max}}.
\end{equation} This is experimentally measured to be $\Pmax = 0.12$.

The average number of photons scattered by a molecule that became dark after interacting with the deflection light is 
\begin{align}
    N &= \sum_{1}^{N_\text{max}}n p (1-p)^{n-1}\\
    &= \frac{1 - \Pmax\left(1 + N_\text{max}\,p\right)}{p}
\end{align}
We experimentally measure $N = 41.1 \pm 6.3$. Combining this with $\Pmax=0.12$, we solve for $N_\text{max}$ and $p$ numerically and find $N_\text{max} = 137.47$ and $p = 0.0153\pm 0.0023$. The uncertainty on $p$ is estimated based on the derivative $\frac{dN}{dp}$ and the uncertainty in the measured $N$. We verify this calculation by using these values in a Monte Carlo trajectory simulation, which takes into account the forward velocity distribution and Poisson statistics for photon scattering, and find a predicted deflection that matches our experimental observations.

% \[\Tilde{X}^2A_1(0_0)[1_{11}] \to \Tilde{A}^2B_2(0_0)[0_{00}]\]